# A Big Data Based Framework for Executing Complex Query Over COVID-19 Datasets (COVID-QF)

***Eman A. Khashan[1*], Ali I. Eldesouky[1], M. Fadel[1] and Sally M. Elghamrawy [2]***

[1] *Department of Computers and Systems, Faculty of Engineering, Mansoura University Mansoura, Egypt;*
[2] *Department of Computer Engineering, MISR Higher Institute for Engineering & Technology, Mansoura, Egypt.*
[1*] *Eng.emankhashan@gmail.com*
[2] *Sally_elghamrawy@ieee.org - IEEE Member*

## *Abstract*

COVID-19's rapid global spread has driven innovative tools for Big Data Analytics. These have guided organizations in all fields of the health industry to track and minimized the effects of virus. Researchers and developers are increasingly required to follow up and detect coronaviruses through artificial intelligence, machine learning, and natural language processing, and to gain a complete understanding of the disease. In Rensselaer Polytechnic Institute (RPI) scientists use large-scale data, analysis and a variety of aspects to better understand coronavirus. Whereas, the corona takes place in huge numbers in the world, with which only big data application and the work of NOSQL databases are suitable. Surely, A system valid for analysing large numbers of data via SQL or large numbers of data via NoSQL is required. The size of the COVID-19 data collected from each country may reach a large volume over time. The SQL form may be insufficient to handle this size, thus in this case the NOSQL databases should be used at certain time. There is a great number of platforms used for processing NOSQL Databases model like: Spark, H2O and Hadoop HDFS/MapReduce, which are proper to control and manage the enormous amount of data and there are great number of platforms used for processing SQL databases models like SQL Server, Oracle, Sybase and MYSQL. Many challenges faced by large applications programmers, especially those that work on the databases of Virus COVID 19 through hybrid data models through different APIs and query. Therefore, a strong, intelligent system needs to be built urgently to save and analyse the produced data. this paper proposed a storage framework able to handle both SQL and NOSQL databases renamed (COVID-QF) for COVID-19 datasets in order to treat and handle the problems caused by virus spreading worldwide clearly by reducing treatment times. In case NOSSQL database Model using Hadoop HDFS/Map Reduce and Apache Spark. The COVID-QF consists of three Layers: data collection layer, storage layer, and query Processing layer. in the data collection layer, the data is collected from various dataset with different size and different type. The storage layer using Hadoop HDFs and MapReduce to divide data into collection of data-saving into processing blocks and spark to connect with Connector for spark to connect with different engine of databases to reduce time of saving and retrieving, while the Processing layer executing the request query and sends results. The proposed framework used three datasets increased for time for COVID-19 data (COVID-19-Merging, COVID-19-inside-Hubei and COVID-19-ex-Hubei) to test experiments of this study. The datasets are used to investigate the validity of COVID-QF framework. The results obtained insure the superiority of the COVID-QF framework.

## 1- Introduction

With the emergence of the emerging corona virus (nCOV-2019), China has started to rely on technologies such as artificial intelligence and big data to limit the spread of the virus in the country. She used the huge database she maintains, with her vast experience in designing collective monitoring tools. How did the big data help China monitor the spread of the Corona virus?

Because telecommunications companies are keen and strict on citizens to register with their real names when ordering telecommunications services, or even buying a new smartphone, the Chinese government has collected a large collection of data about its citizens across the country, using it to build tools that enable it to easily track people who travelled in the period The last one to the Chinese city of Wuhan where the virus appeared. Coinciding with the spread of the Corona virus, the state-owned China Electronics Group launched a new application called (Close Contact Detector), also in cooperation with many Chinese institutions such as: State Council, the National Health Committee, the Ministry of Transport and Railways, and the China Aviation Authority. Where the principle of the application's work depends on that any citizen can register using the phone number, and then he must enter his name and ID number. Its data will be matched with the large database of public authorities to help people know if they have been in close contact with anyone who has had the virus in the past two weeks. The data is mainly derived from the China Railways Corporation database. Which has records in from the big data on Chinese citizens who have been using trains for 20 years. That is why this study can say that the big data have helped China in a big way to quickly count the numbers and thus limit the disease. The data collected from China or from all parts of the world is often from different types of SQL or NOSQL databases. Whereas, the corona takes place in huge numbers in the world, with which only big data application and the work of NOSQL databases are suitable. The proposed system is valid for analysing large numbers of data via SQL or large numbers of data via NOSQL.

The probability of a great deal of concern the popularity of NoSQL systems is caused by their efficiency in handling unstructured data and backing up effective design schemes that give the system users supreme flexibility and scalability. This paper identifies a relational database and several categories of NOSQL Databases with structural features: key-value, graph, column and document databases. Likewise, every NoSQL database has a special query language and does not support the criteria of other systems. The main problem that many researches focused on, is that there is no standard way for expressing, executing and optimizing complex queries across NOSQL Databases [1,4]. Currently, data stores have several diversified APIs. The programmers of applications based on multiple data stores must be familiar with these APIs during the process of coding these applications. As a result of the variety and changes in the data models of various databases, there is no standard way to solve the problem of implementing queries for various NoSQL data stores. The reason is due to a lack of a combined access model for diversified data stores. The programmers must challenge themselves with the execution of these queries, which are hard to optimize. On the other hand, optimization puts certain criteria into consideration, such as data transformation and movement costs, which might be expensive for big data [31].

All of these reasons encourage sharing in the interoperability between two or more varied and powerful frameworks. In this paper, Mongo and Cassandra focus on being the most popular way to help companies make business decisions. Several researchers and developers have focused on this problem. The variety of relational and NoSQL data models (relational, key value, ordered key-value, document, semi-structured and graph databases) and query languages (SQL, Cassandra Query Language (CQL), MapReduce querying language, etc.) is the main difficulty. Salami et al. [1] Identify a common data model and use algebra to address complex declarative inquiries. In this technique, queries are handled in multiple data stores called VDS (Virtual Data Store), that is, default data stores. The optimization stage is carried out by a two-step broker. First, the selection and project processes are defined down to the local data stores. This allows to reduce the amount of data exchange. Second, an ideal distributed plan is designed with a dynamic programming method. The distributed plan seeks to reduce I / O and CPU costs and to charge and convert data. However, this technique is limited to redressing an ODBAPI query and some query operators. Another method was developed by P.Sangat et al. [2] called DIMS. In DIMS Most data generated by ubiquitous sensing applications have the character of time series, such as monitoring data of power station, and from others a pattern of interrelationship emerges, for instance the correlation between patients, disease, and symptoms. Further, high sampling frequency and high data generation rate also feature. To satisfy the needs of various requirements, a data storage system should have various abilities, such as making different schemes and profiles for different applications. Song et al [15], they present the design, implementation and evaluation of Haery, a column dedicated to big data. Haery is built on Hadoop HDFS and distributed computing framework relying on MapReduce. Haery's download and query performance results are the most stable and effective. But there is more cost in time when data volume increases. Haery proposed the following models and algorithms: Key-Cube, an improved Z-order based linearization algorithm and an address tree, Accumulation, which is a key-cube expansion approach, Query algorithms to implement queries on key-cubes and physical storage and the system architecture, components and implementation of Haery. The rest of this paper is organized as follows. In section 2, this paper presents the related work. In section 3, the proposed COVID-QF framework, which has three layers. In section 4, research work discusses implementation and evaluation of COVID-QF. Section 5 provides conclusion and future work.

## 2- Related Work

Several researchers and developers have focused on this problem. The variety of relational and NoSQL data models (relational, key value, ordered key-value, document, semi-structured and graph databases) and query languages (SQL, Cassandra Query Language (CQL), MapReduce querying language, etc.) is the main difficulty. G. Baruffa et al. [3] characterized a Spectrum Sensing that provides a service which allow end users to easily access and process wireless spectrum data. To reduce the latency of services provided by the platform, that adjust the data processing chain, they took an interest in Mongo and Cassandra databases and did not consider the rest of the databases. Khan et al. [4]. and Duggan et al. [5] it offers frameworks that called PolyWeb and BigDAWG, respectively. PolyWeb and BigDAWG retain data sources in a primary format, that is, without serializing them in a common data format. In PolyWeb, SPARQL queries are translated into the original query language for these sources. PolyWeb indexes each data source to predict the query and creates deep left plans. Despite the efficiency, the current methods are not able to exploit knowledge about the main features of integrated data sources, and produce custom query plans for selected sources to collect data from the data lake. In contrast, the QODM [6] approach produces distinct schema using the data model and data schema of an application for NOSQL Databases. This approach will not prevent programmers from using any NoSQL database. Document and relational data stores are integrated in a hybrid mediation approach proposed by Roijackers et al. [7]. However, this approach does not consider other NoSQL data stores. Tsimmis et al [10]. presented a mediation platform based on rewriting queries. A semi-structured data model called OEM was suggested by Tsimmis and grants support for several data stores using the global schema and related query language (Lorel). In Lorel, a global schema is used to rewrite queries, and this method is considered a view approach for the data sources, but lacks query optimization. Sharma et al. [8] are studying the performance of RDBMS, Document based No SQL data base (MongoDB) and Graph based No SQL Data base (Neo4j). they got unexpected results in case of neo4j as it took longest time as compared to MongoDB and PostGre SQL. IBM NoSQL [11] is a commercial solution that permits a database to store relational and NoSQL data in the same data store. A problem is that this solution does not support accessing the database from outside of IBM servers. S. K Pandey et al. [12] presented the CBCQL framework, which is internally mapped to CQL and so has the same power as Cassandra, but this framework does not support other NOSQL Databases. A relational database is moved to Apache Cassandra by designing a data model of the application designed by the MySQL database by Aaron Schram and Kenneth M. Anderson [13]. This design does not explain the process of implementation for other applications. However, few NOSQL Databases are supported only by these frameworks, so the programmer has to make designs for data models of an application and choose a proper strategy for data mapping. COVID-QF is a proposed framework for improving and estimating complex queries for relational databases and other types of NoSQL data stores. For this purpose, a unified data model is proposed that uses a suitable environment such as Apache Spark with MongoDB [18,24] to optimize the qualification of the data ingestion process. The COVID-QF framework transforms each query process received from any dataset to the matched Engine after using Hadoop/HDFS and Hadoop/MapReduce with parallel k-means clustering for processing data without physical transformation data.

## 3- Proposed COVID-QF Framework

COVID-QF proposes to increase both Corona Virus dataset (COVID-19) injection and query efficiency. s. COVID QF contains three entry, index, inventory and query phases. This section introduces the proposed COVID-QF approach, which is capable of executing complex queries across various datasets. COVID-QF uses the internal architecture of the Apache Spark engine to enhance processing performance and reduce computer time. Apache Spark [9] is an internal memory distributed data processing system, a data engine which performs tasks as quickly as 100 times as multiple step tasks This framework consists of three layers, as shown in Figure 1: Collecting data from various datasets layer, Storage layer, and query processing layer. In the following sections, this paper discusses the different layers of the proposed COVID-QF framework.

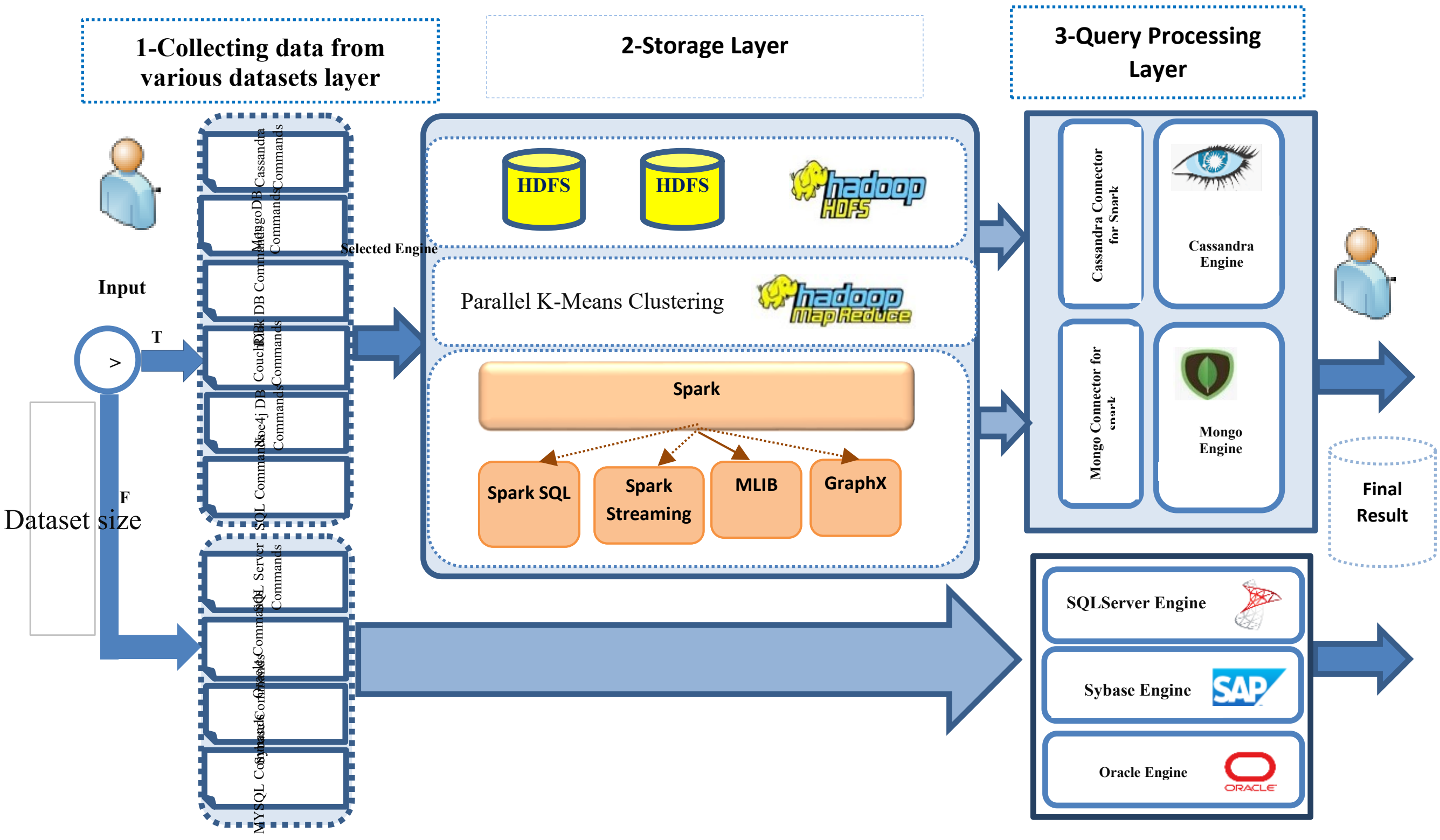

**Figure 1**. COVID-QF Framework

### *3-1 Collecting data from various datasets layer*

This layer receives any SQL or NoSQL dataset query to match the sentences of the query given by the user with the stored libraries that hold a number of statements for each database type either SQL or NoSQL from the database engine and then compares the sentence with the stored libraries to define the required database engine. This paper prepared a set of libraries for each of the databases that are studied, such as SQL as example of relational database and MongoDB, Cassandra as an example of NOSQL Databases. Indeed, this approach symbolizes the combined parts among every deployed data storage and delivers a unified model to the following layer of the framework. This model contains the particular operations of every database. It is noteworthy that the user has to add a particular implementation of the data store if he/she needs to integrate an extra database. In the following figures, an explanation is given for testing the query statements for the databases used. This paper used SQL Server (as an example of a relational database), MongoDB and Cassandra (as examples of NOSQL Databases). Figure 2a explains the stored SQL libraries statements for SQL database while Figures 2b, 2c,2d,2e and 2f explain the CRUD statements for MongoDB, Cassandra, Riak, Couch and NOE4J DB, respectively, as examples of the NOSQL Database libraries used in this paper.

**Select Operation**

**Select * from People;**

**Select * from People where id=3**

**Insert Operation**

Insert into People(id, Name, position, phone) values(1,’John’,’Egypt’,’0100000’)

**Update Operation**

Update People set position=’USA’ where id=25

**Delete Operation**

Delete * from People where id=25

**Figure 2a**. SQL libraries.

**Select Operation**

- db. People. Find ();

**Insert Operation**

- db. People. Insert ({cust_id: 'appl01', branch: 'main', status: ‘A’})

**Update Operation**

- db. People. Update ({custage: {$gt: 2}}, {$set: {branch: 'main’}}, {multi: true})

**Delete Operation**

- db. PeopleCollection.deletemany();
- db. PeopleCollection.remove();

**Figure 2b**. Mongo libraries.

**Select Operation**

- SELECT * FROM People;

**Insert Operation**

- INSERT INTO People (custid, branch, status) VALUES ('appl01', 'main', 'A');

**Update Operation**

- UPDATE People SET comments ='='Rides hard, gets along with others, a real winner' WHERE id = fb372533-eb95-4bb4-8685-6ef6 1e994caa IF EXISTS;

**Delete Operation**

- DELETE lastname FROM People WHERE id = ‘c7fceba0-c141-4207-9494-a29f98de6f’;
- DELETE FROM DB. People WHERE id= 2;

**Figure 2c.** Cassandra libraries

**Select Operation**

riak-shell>select Name, Position from People where id > 1234560 and region = 'South Atlantic' and state = 'South Carolina'

**Insert Operation**

riak-shell>INSERT INTO People VALUES ('SC', '2018-01-01T15:00:00', 'sunny', 43.2, 0x3af6240c1000035dbc), ('SC', '2017-01-01T16:00:00', 'cloudy', 41.5, 0x3af557bc4000042dbc), ('SC', '2017-01-01T17:00:00', 'windy', 33.0, 0x3af002ee10000a2dbc);

**Update Operation**

riak-shell>Update People set position=’USA’ where id=25

**Delete Operation**

riak-shell>Delete * from People where id=25

**Figure 2d**. Riak libraries.

**Select Operation**
```
"selector": {
    "year": {"$gt": 2010}   },
  "fields": ["_id", "_rev", "year", "title"],"sort": [{"year": "asc"}],
  "limit": 2,   "skip": 0,
  "execution_stats": true
```
**Insert Operation**
```
INSERT INTO `travel-sample` (KEY, VALUE) VALUES ("key1", { "type" : "hotel", "name" : "new hotel" }) RETURNING *
```
**Update Operation**
```
curl -X PUT http://127.0.0.1:5984/database_name/document_id/ -d '{"field" : "value", "_rev" : "revision id" }'
```
**Delete Operation**
```
$ curl -X DELETE http://127.0.0.1:5984/my_database/001?rev=1-
3fcc78daac7a90803f0a5e383
{"ok":true,"id":"001","rev":"2a561d56de1ce3305d693bd156"}
```

**Figure 2e**. Couch libraries.

**Select Operation**
```
MATCH (C:People)
WHERE 3 <= p.yearsExp <= 7
RETURN p
```
**Insert Operation**

Creating a node:
```
$ CREATE (n)
```
Creating a Node with a Label:
- $ CREATE (node1: Test

Creating multiple Nodes with unique Labels simultaneously:
- $ CREATE (node1: test), (node2: Test2), (node: Test3)

Creating Nodes with Properties:
- $ CREATE (node1:Test {nodeId: 2, nodeName: 'sample', nodeDescription: 'testing'})return node1
- Setting Properties when creating:
- $ CREATE (node1: Test) set node1.name-'test' return node1

Creating a Relationship:
- $ CREATE (emp:Employee), (pro:Project) ,(emp)-(ew:EMP_WORKS_FOR_PRO)->(pro) return emp, pro

**Update Operation**
```
MATCH (mgr:People {PeopleID:5})
MATCH (cust: People { PeopleID:3})-[rel:REPORTS_TO]->()
DELETE rel
CREATE (cust)-[:REPORTS_TO]->(mgr)
RETURN *;
```
**Delete Operation**
```
MATCH (mgr:People {PeopleID:5})
MATCH (cust: People { PeopleID:3})-[rel:REPORTS_TO]->()
DELETE rel
```

**Figure 2f**. NOE4J libraries.

Algorithm 1 illustrates the method of discovering the database type of the query to be executed based on the libraries stored in the application to show the selected engine database.

```
Input: qs query Statement
Output: SQL or NoSQL database Engine
  1.   parsing Query Statement (qs).
  2.   Declare arr[6]={ sql, Mongo, Cassandra, riak, Neo4j, Couch}
  3.   For i =0 to 5
       {
       if(arr[i]  =  qs)  selected_Engine=arr[i]
       break
       }
  4.   Switch selected_Engine
       Case sql
              Connect to sql Engine
       Case Mongo
              Connect to Mongo Engine
       Case Cassandra
              Connect to Cassandra Engine
       Case riak
              Connect to Riak Engine
       Case Neo4j
              Connect to Neo4j Engine
       Case Couch
              Connect to Couch Engine
       Case else
              No Engine
       End switch
  5.   If selected_Engine='Noengine', then display error page and stop running,
       else continue running & execute query qs.
```

**Algorithm** 1. Matching Selector algorithm

According to **Algorithm** 1, the results of matching patterns and input values, one of the following decisions will be followed:
If the patterns are identical to the SQL database, the application will continue to run the path of the SQL database. If the patterns are identical to the Mongo database, the application will continue to run the path of the Mongo database. If the patterns are appropriate for the Cassandra database, the path for the Cassandra database will be followed.

If you want to apply patterns to other databases, you must add their own libraries

### *3-2 COVID-QF Storage Layer*

This paper mentioned previously that the volume of data received from various sources to monitor the infected with COVID-19 virus is very huge. This section introduces the second layer of COVID-QF framework which responsible for storing the collecting data from various resources. COVID-QF deployed and used Hadoop/HDFS [1] to store the incoming data. Hadoop is an open source distributed computing platform that mainly consists of the distributed computing framework MapReduce and the distributed document system HDFS [3]. The formula (1) uses to calculate HDFS node storage (*H*) required:

*H*: denoted the HDFS node storage required
*C*: is the compression ratio and completely depends on the type of compression used and size of the data.
*R*: It is the replication factor which is 3 by default in production cluster.
*S*: S denotes the initial amount of data you need to move to Hadoop.
*I*: I represent the intermediate data factor which is usually 1/3 or ¼. It is Hadoop's intermediate working space used to store the intermediate results of different tools like Hive, Pig etc

1.2: 1.2 or 120% more than the total size.

$$H = \frac{C*R*S}{(1-I)*1.2} \qquad (1)$$

MapReduce [9,22] is a software platform for parallel processing programming of large-scale data pieces. The MapReduce strategy is applied to the k-means clustering algorithm and clustered for the data factors. The k-means [19] algorithm can be successfully parallelized and clustered on hardware resources. MapReduce can be utilized for k-means clustering. The results also show that the clusters shaped using MapReduce are similar to the clusters produced using a sequential algorithm. Once HDFS takes data, this process breaks information down into separate blocks and distributes those blocks to different nodes in the cluster, thus enabling high-efficiency parallel processing. The data from HDFS is accessed by a Spark streaming program for handling before being stored in MongoDB in the server of the database. Resilient distributed datasets (RDDs) are an abstraction presented by Spark [13]. RDDs symbolize a read-only multiset of data objects divided into a group of machines that continue operating as designed despite internal or external changes (fault-tolerant way). Spark is considered the first system of programming languages in general and is used as an interactive way to handle big data sets for COVID-19 from each country in the world.

### *3-3 Query Processing Layer*

Instead of storing the COVID-19 data as tables with columns and rows, the data are stored as documents. Every document can be one of the relational matrices of the numerical values or the overlapping interrelated arrays or matrices. These documents are serialized as JSON objects and stored internally using JSON binary encryption known as BSON in MongoDB; the data is partitioned and stored on several servers called shard servers for simultaneous access and effective read/write operations. MongoDB and Apache Spark are integrated seamlessly by this connector. MongoDB aggregation pipelines and a problem of how to assign a group of objects into groups, called blocks, so that the objects within the same group, partitioning is by using a cluster assignment function $C: X \rightarrow \{1,2, \ldots., k\}$ when X is a set of objects, the Number of clusters $K \in \mathbb{Z}^{+}$ and Distance function $d \in \mathcal{R}0^{+}$ between all pairs of objects in X, partition $X$ into $K$ disjoint sets $x1, x2, \ldots.., x_k$ such that $\sum_k \sum_{x,x' \in X_k} d(x, x')$ With N = |X|, the number of distinct cluster assignments possible as follows [33]:

$$S(N, K) = \frac{1}{K!} \sum_{k=1}^{K} -1^{K-k} \binom{K}{k} k^{N} \qquad (2)$$

#### *3-3-1 MongoDB Engine*

Sharding is a way to distribute data across multiple devices. This paper presents MongoDB, which uses sharding to support deployments using very large datasets and high-productivity processes. Database systems that contain large datasets or high-productivity applications can challenge the capacity of a single server. For example, high query rates can exhaust the CPU capacity for the server. A range of sizes greater than the system's RAM can help to confirm the I/O capacity of the drivers. A database can have a mixture of sharded and unsharded collections. Sharded collections are partitioned and distributed across the shards in a cluster. Unsharded collections are stored on a primary shard. Each database has its own primary shard, as shown in Figure 3.

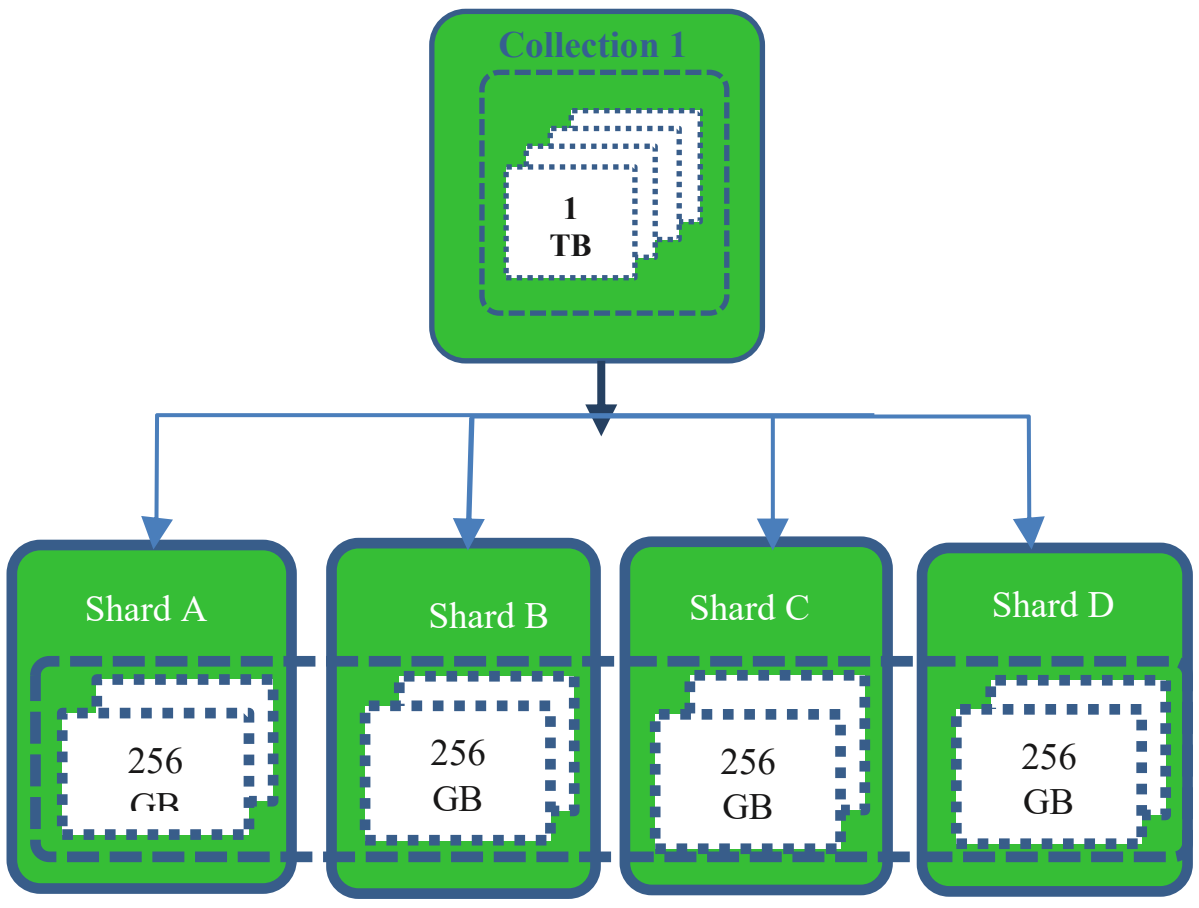

**Figure 3.** Sharding Mongo DB Stage

The Mongo DB uses the following equations to measure the theoretical maximum collection size. This Study supposes *M* is the max Splits, *md* The maximum BSON document size is 16MB or 16777216 bytes. mb is the maximum collection size(mb), *C* is the chunk size, and *avg* is the average size of shard key values in bytes.

$$M = \frac{md}{<avg>} \quad (3)$$

$$MB = M * \frac{C}{2} \quad (4)$$

### *3-3-2 Cassandra Engine*

The Apache Cassandra database has linear scalability and proven tolerance for hardware or cloud infrastructure, and these attributes make this database an ideal platform for important data. This paper presents replication supported by the Cassandra database across multiple data centres that is best in class, providing less downtime for users and peace of mind by knowing that it can overcome regional interruptions. This paper proposes two kinds of partitioning methods that can work with the Cassandra database: vertex partitioning and edge partitioning. Later, this study will introduce how can research paper dealing with these methods. This paper investigates vertex partitioning and edge partitioning to show differences in the results about them. This paper investigates vertex partitioning and edge partitioning to show differences in the results about them.

### *3-3-3 Data Partitioning*

Cassandra divides the database into smaller, partially overlapping datasets that are stored locally on each node. Thus, unlike other NOSQL Databases such as HBase, Cassandra does not require a shared file system (for example, HDFS). A hash function is used to distribute basic registry keys for the nodes. This process is performed by dividing the scope of the hash key into subdomains called partitions (also called token ranges). In blocks without repeating (RF = 1), each node can be configured to store unique partitions locally. In this section, the necessary background will be provided and presented with the data and account models we target. Table 1 contains the partitions variables used in this paper. This paper uses formula (5) to calculate the size of data partitions.[35]

$$NV = Nr(Nc - NpK - Ns) + Ns \quad (5)$$

In order to determine the size, this study uses formula (6) to determine the size St of a partition [35]:

$$St = \sum_i SizeOf(Ck_i) + \sum_j SizeOf(Cs_j) + Nr \times (\sum_k SizeOf(Cr_k) + \sum_L SizeOf(Cc_l)) + SizeOf(t_{avg}) \times Nv \quad (6)$$

**Table 1:** The partitions variables used in this paper

| Symbol | Description |
|---|---|
| *Nv* | The number of values (or cells) in the partition |
| *Nr* | The number of values per row. |
| *Nc* | the number of columns. |
| *NpK* | the number of primary key columns. |
| *Ns* | The static columns. |
| *St* | The size of partitions. |
| *Ck* | partition key columns. |
| *Cs* | static columns. |
| *Cr* | regular columns. |
| *Cc* | clustering columns. |
| *tavg* | the average number of bytes of metadata stored per cell, such as timestamps. |
| *sizeOf()* | This function refers to the size in bytes of the CQL data type of each referenced column. |

## 4- COVID-QF Experiments

COVID-QF is utilized to store, manage and execute queries of big data and greatly facilitates the developer's task. In this paper, the proposed model rewrites each query into the particular query language of the integration data store. The processing layer in COVID-QF turns results into a suitable format such as JSON before responding to the system users. Therefore, the overhead is considered reasonable to some extent. Because of memory management trouble in the driver, there is a probability that the performance of COVID-QF will degrade after 50000 entities. The results of experiments testing MongoDB and Cassandra DB are shown in the following sections.

### *4-1 Datasets*

This paper utilized a real coronavirus infected people's dataset. Three databases of varying sizes are included. This was called COVID-19-Merging [33] for the first time. This dataset is generated by combining data sets from various countries to generate large-scale datasets. COVID-19-inside-Hubei [34], a second dataset comprising of people living with the virus, has 27,870 patients inside the Chinese area of Hubei [35]. The third dataset, called COVID-19-ex-Hubei [33], comprising of contaminated persons, has 160,175 records for patients outside of the Hubei Province of China.

### *4-2 System analysis details*

This section provides brief information on the techniques used and preparing the environment for the experiment. This study creates and publishes Hadoop, HDFS, Apache Spark, MongoDB, MongoDB link for Spark, Cassandra DB connector, Cassandra DB for spark server and SQL on the environment with the following specifications.

1. Hadoop / HDFS Setup: This paper deploys Hadoop / HDFS version 2.10.0 using standard configuration parameters. The application server is running on HDFS to store incoming data and access software Spark Access data from HDFS for processing before inclusion in SQL, MongoDB and Cassandra DB in the database server.
2. Spark Setting: Apache Spark is a widely open source framework. Spark introduces a stripping called elastic Distributed Data Sets (RDDs), which represent a multiple read-only set of data items divided across a set of devices that are maintained in Error tolerant method. Spark is the first system to interactively use a general-purpose programming language to collaboratively Large data sets on a block. Apache Spark 2.3.2 version is published in standalone mode and control configs of application code. For example, the experiments have set spark.serializer as *org.apache.spark.serializer.KryoSerializer*. The application was created Using Scala 2.10.4.

3. Database Connector for Spark: This connector provides a seamless integration between matched Database and Apache Spark. It effectively uses database assembly lines and secondary indexes to extract, filter and process the sub-data required for the Spark process. Additionally, to maximize performance over a large distribute Datasets, they link the RDDs to the source database node and reduce the data transfer across the cluster.4. Hardware: The Spark app server has 16 dedicated hubs, 64GB of memory, 459GB of hard drives, and 64-bit Ubuntu GNU / Linux. The mongo dB database server and both Shard 4 server have dedicated cores, 16GB memory and 130GB HDD, while each of the initialization servers contains 1 hard disk, 4GB and 30GB. The Cassandra database server. 16 GB memory, 130 GB hard drives, and Microsoft SQL 2017 server has been installed with the same infrastructure specifications previously mentioned.

### *4-3 Results Analysis and Discussion*

In this section, the core algorithms of COVID-QF and the results are evaluated, and compared with some popular NoSQL and relational databases using generated datasets and various query workload. The experiments conducted provided a comparison between COVID-QF and ODBPI to measure the cost time based on the number of joins. The results obtained, when using the Hadoop and spark, reflects higher performance compared to recent algorithms. COVID-QF evaluation implemented two types of joins: linear join and star join. And the results obtained from linear joins proved that they are better than star joins results. In addition, a comparison with the DIMS framework to measure the average time and ingestion time have been implemented using two different databases: Mongo and Cassandra databases. The COVID-QF results when using Mongo with sharding technique is better than using Mongo without sharding technique, especially when the size of the data is very large. When the comparison was done on the Cassandra database, it got better results when using the portioning technique than using it without portioning. This study also compared two types of partitioning in Cassandra database. According to COVID-QF experiments the edge portioning has got better results than vertex, especially when using a large size of data. When adding the comparison process with Haery framework, the Cassandra with COVID-QF framework achieved better results than Haery, but the results of Haery using Mongo database are relatively better than COVID-QF results with no sharding. On the other hand, when COVID-QF applied the sharding technique, the results obtained are better than results obtained from Mongo and Cassandra databases, when using a large size of data.

### *4-4 Cost Model*

The cost of implementation is the sum of the costs of each process that composes the implementation plan. It should be noted that the cost does not directly represent time. Of course, more cost means more time. It is used to compare two query execution plans, but not to directly estimate response time. To evaluate the cost formula, the matrix multiplication between the row vector containing the coefficients $\alpha$, $\beta$, and $\gamma$ was calculated. A column vector contains the values of the parameters defined in the catalog, and a fixed variable called const which is a scalar and can be a cardinality, selectivity, etc. In addition, if the parameter does not depend on a specific measurement (CPU cost, I/O cost, or cost of connections), this will take the latter an empty value in the column vector. Matrix multiplication [1] is calculated as follows:

$$(const, \alpha, \beta, \gamma)\begin{pmatrix} t_{cpu} \\ t_{i/o} \\ t_{cconn} \end{pmatrix} = const\left(\alpha \times t_{cpu+}\beta \times t_{i/_{o}+}\gamma \times t_{cconn}\right) \quad (7)$$

Performance estimation is an important point for a new framework. This estimation is shown in the outcomes of total cost, average time, and ingestion rate. These outcomes are utilized to estimate the efficiency of the proposed framework. The outcomes are calculated by the following equations:

$$TotalCost = \alpha \times t_{cpu+}\beta \times t_{I/o+}\gamma \times t_{cconn} \quad (8)$$

$$Average\ time = \frac{Total\ cost}{No.of\ joins} \quad (9)$$

$$Ingestion\ rate = \frac{No.of\ records}{Average\ time} \quad (10)$$

### *4-4-1 Data Ingestion*

In this section, the time required for suggesting a COVID-QF will be calculated utilizing another data set. COVID-19-Merging [33], COVID-19-in-Hubei [34] and COVID-19-out-Hubei [35] as mentioned above are used in three separate datasets. The following sets are used in various sizes of data. The cumulative time required to index a particular data set is shown in Figure 4. As shown in the chart, different data sets require some period. The time needed is through slowly as the data collection grows. It takes approximately 340 Ms for the COVID-19 combining databases to save and index more than one million documents. This research paper has performed data archiving from the various databases mentioned previously. Three types of databases were also used, the SQL database, and also the Mongo database, as well as the Cassandra as examples of NOSQL databases. It was noted that the Cassandra database, followed by Mongo, carried out the process of storing data faster than relational databases.

The research paper worked on calculating the total time calculated in the process of saving millions of data in the Three datasets mentioned above using Cassandra data store. The proposed system (COVID-QF) achieved the least time in the process of saving data when making a comparison between two previous systems, the first ODBAPI, which suggested the creation of Virtual Data Store (VDS) as a mediator to executing the user queries in the cloud environment. As for the second framework DIMS, which was working on calculating storing time, ingestion time and retrieving time in mongo Database using Apache Hadoop and spark connector for Mongo and in spark as shown in figure 5.

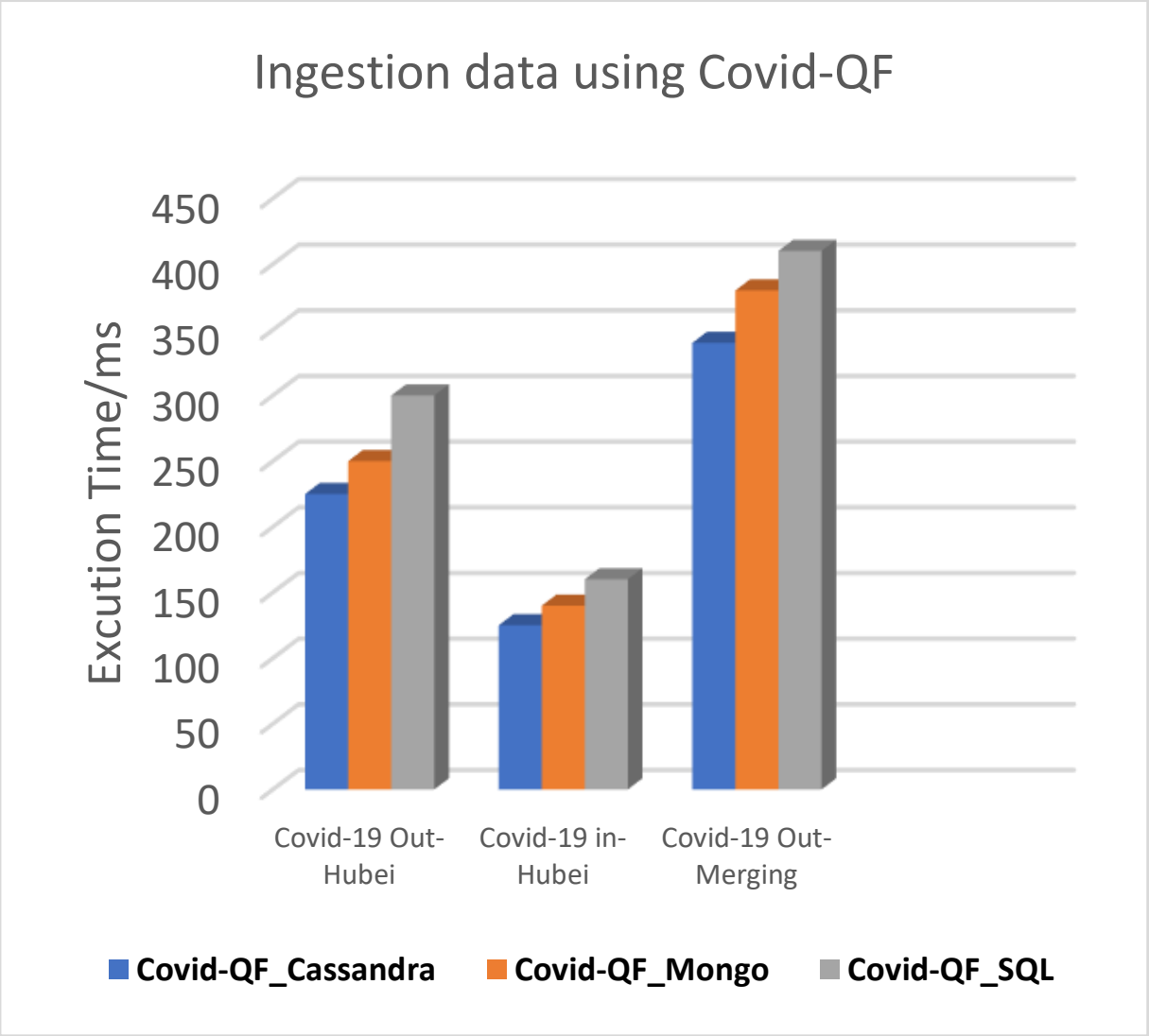

**Figure. 4:** Comparison between Cassandra, Mongo and SQL for executing Total Time to insert data using Covid-QF.

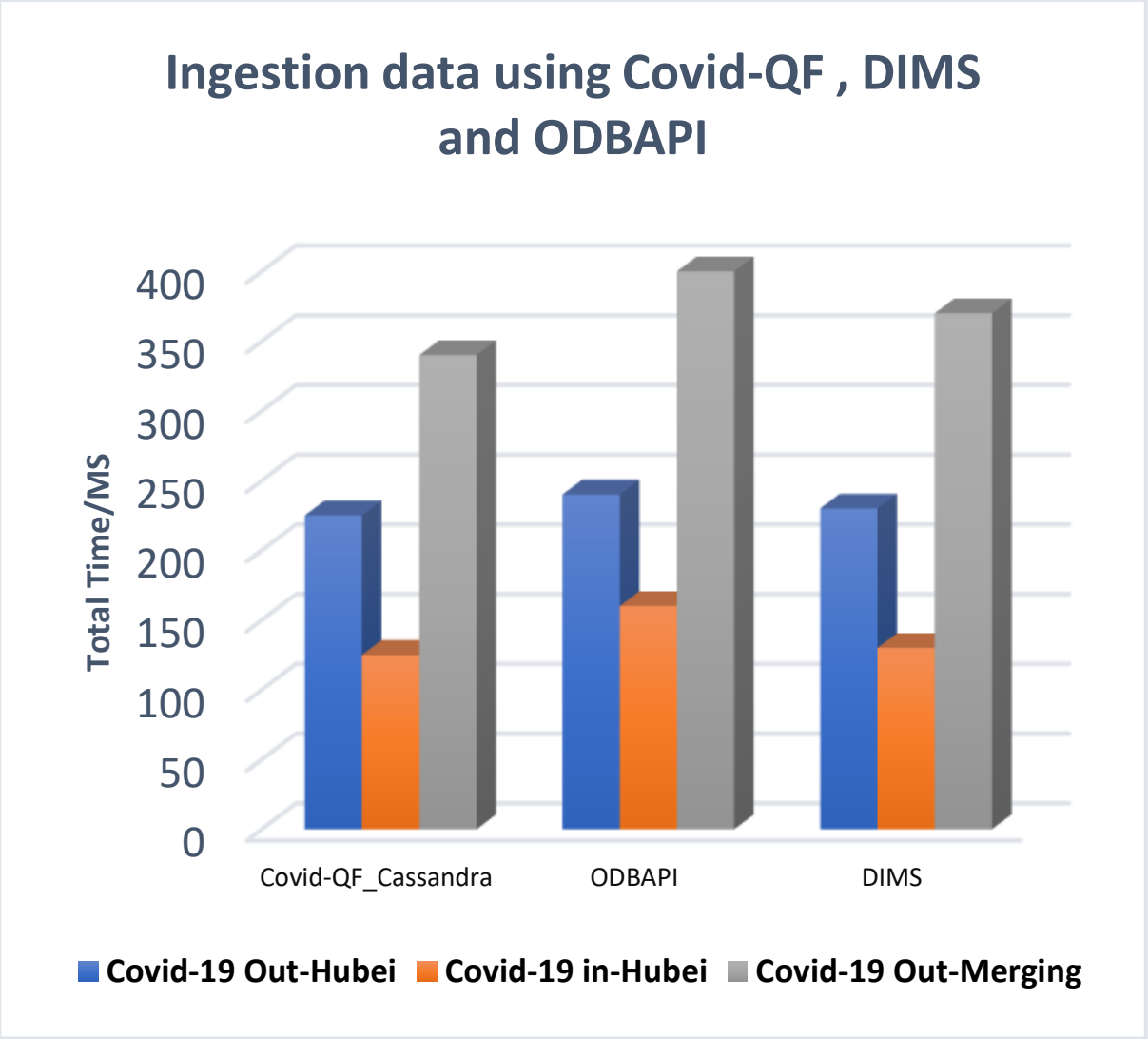

**Figure. 5:** Comparison between COVID-QF, ODBAPI and DIMS for executing Total Time to insert data.

### *4-4-2 Total time for Retrieving data*

The method to scan, pick and retrieve correct data from a database is the data recovery procedure. The database scans for and retrieves the necessary data on the basis of the available queries or orders. Applications will create data in different formats, store it in a register, print it or show it. Parallel processors are also used to speed up data recovery [27] Using MongoDB, semi-structured data (JSON) used in the analysis can be processed and constant adjustments in data which can take place over time can be accommodated. The related data and query language for this experiment is given by MongoDB. This study certain problems however that cannot be dealt with just by MongoDB. Next, to uncover data patterns and developments, we will carry out predictive data extraction. The experiments with MongoDB are complicated but are typically carried out as MapReduce programs. Secondly, all MongoDB queries have a single domain category. Data cannot however be grouped together in separate classes to derive useful information from the results. The usage of large data management systems, including the Apache Spark, to collect data is an option to the MongoDB query language. We may have used other large data analysis systems such as Apache Hadoop, but Apache Spark was superior to Hadoop in carrying out analytical challenge [16,28]. In this section, the time required for suggesting a COVID-QF will be calculated utilizing another data set. COVID-19-Merging [33], COVID-19-in-Hubei [34] and COVID-19-out-Hubei [35] as mentioned above are used in three separate datasets. The following sets are used in various sizes of data. The cumulative time required to index a particular data set is shown in Figure 6. As shown in the chart, different data sets require some period. The time needed is through slowly as the data retrieval grows. It takes approximately 360 Ms for the COVID-19 combining databases to select more than one million documents. This research paper has performed data archiving from the various databases mentioned previously. Three types of databases were also used, the SQL database, and also the Mongo database, as well as the Cassandra as examples of NOSQL databases. It was noted that the Cassandra database, followed by Mongo, carried out the process of storing data faster than relational databases. Also, this research paper worked on calculating the total time calculated in the process of retrieving data in the Three datasets mentioned above. The proposed system (COVID-QF) achieved the least time in the process of retrieving data when making a comparison between two previous systems, the first ODBAPI, which suggested the creation of Virtual Data Store (VDS) as a mediator to executing the user queries in the cloud environment. As for the second framework DIMS, which was working on calculating storing time, ingestion time and retrieving time in mongo Database using Apache Hadoop and spark connector for Mongo and in spark as shown in figure 7.

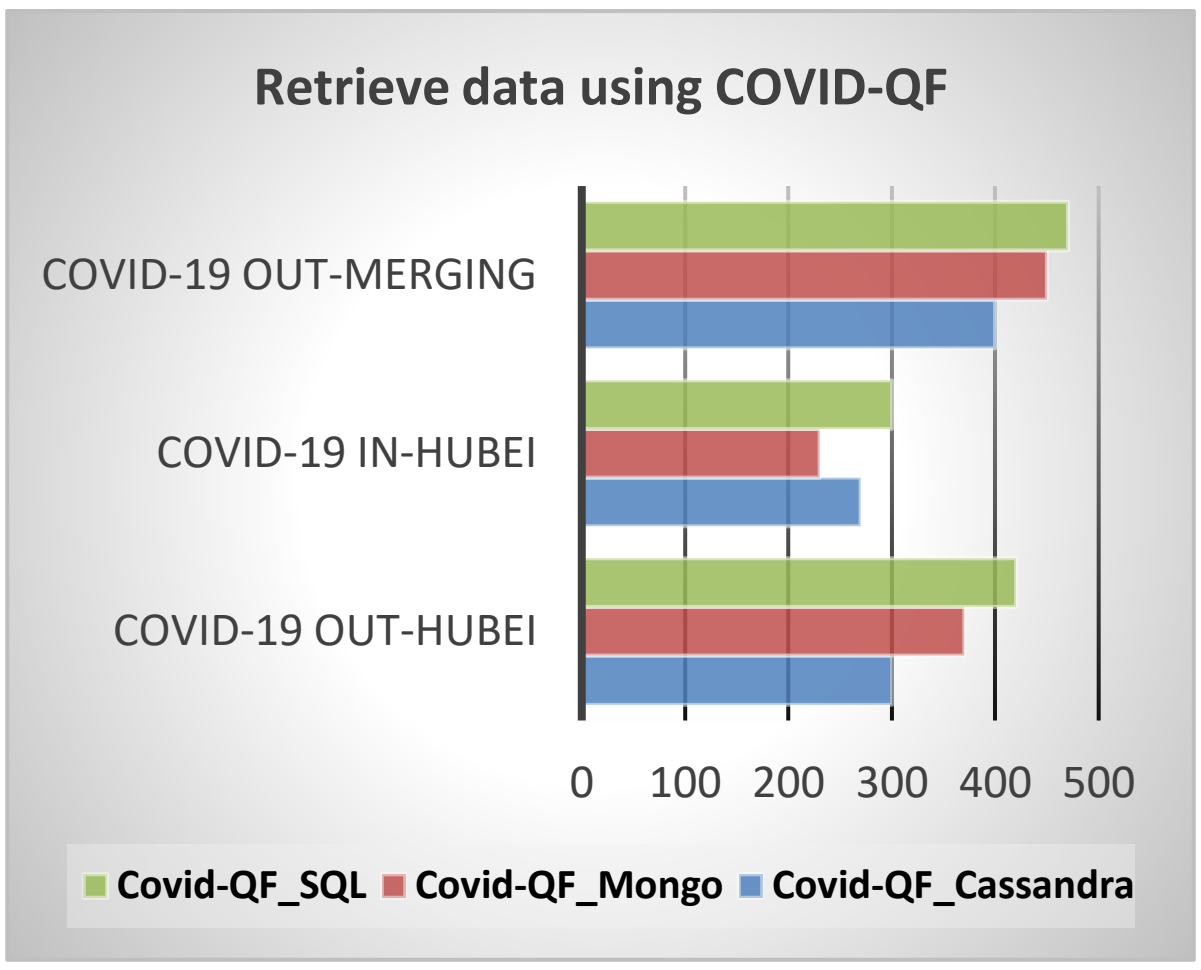

**Figure. 6:** Comparison between Cassandra, Mongo and SQL for executing Time to retrieve data using Covid-QF framework.

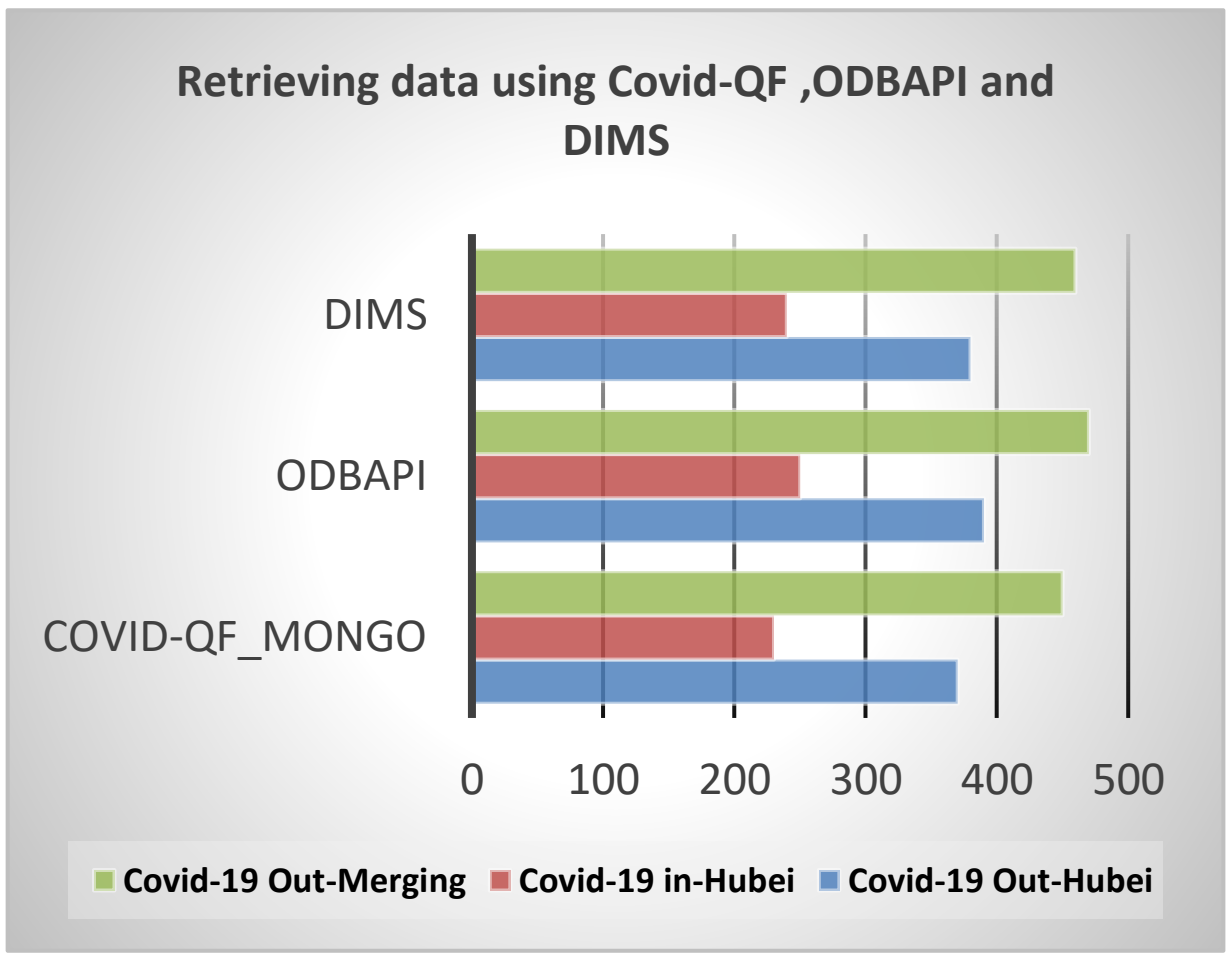

**Figure. 7:** Comparison between COVID-QF, ODBAPI and DIMS for executing Time to Retrieve data.

### *4-4-3 Cost Time for Join Querying Process*

In this Experiment, this study tests the interface querying phase for many datasets of various sizes in this experiment. COVID-19-Merging [33], COVID-19-inside-Hubei [34] and COVID-19-outside-Hubei [35] are used for three separate sets of results. The Join_No=3, Join_No = 5, Join_No = 7 and Join_No = 9 are used for each data collection. Per tuple of separate no of joins shows in figure 8 the Cost Time. As we may see, the time spent utilizing less than the smaller number of Joins is that the necessary amount of k. The time needed for Join_No = 9 is not significant, for example, then that necessary for Join_No = 3.

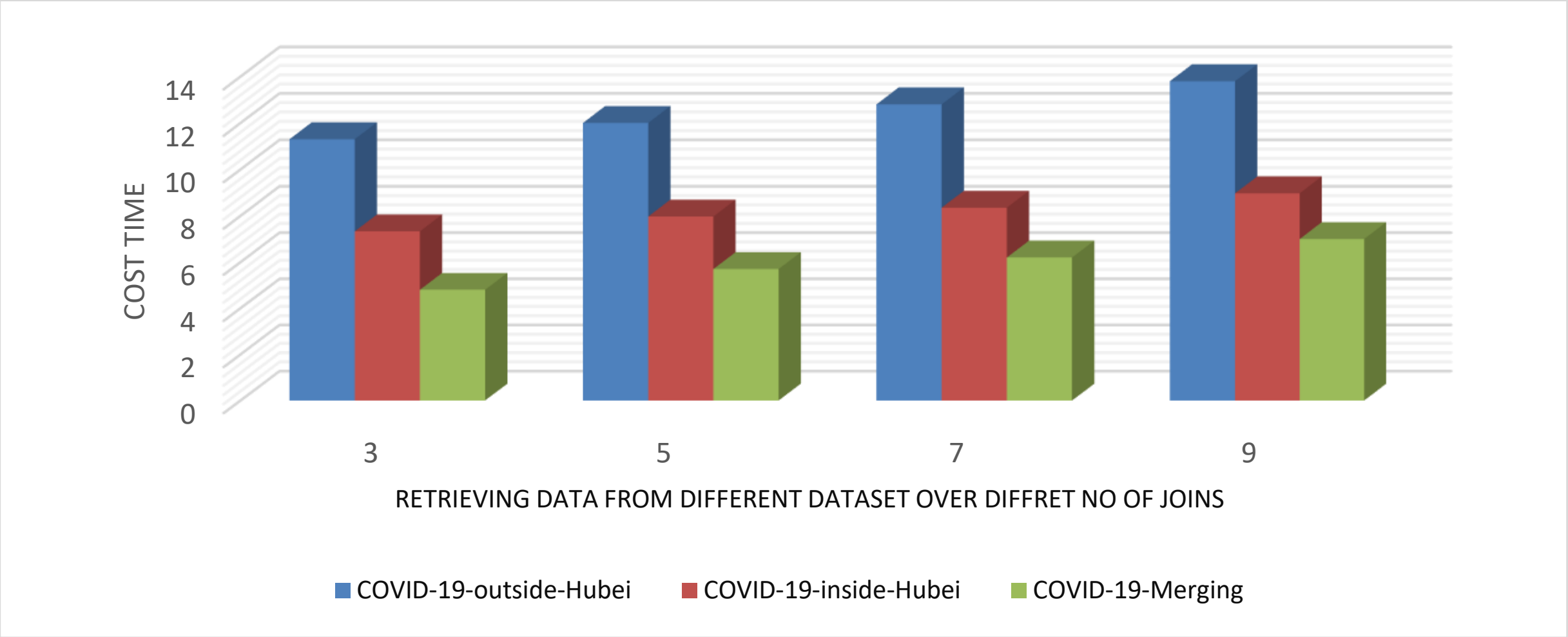

**Figure. 8:** retrieving Covid-19 data from different dataset over different No of Joins

### *4-4-4 Retrieving COVID-19 Data Statistics*

In this section there is a set of statistics that were derived by the proposed framework COVID-QF. In Figure 9 shows the Retrieving Total CORONA Virus Cases Using COVID-QF framework. When Figure 10 shows Active Corona Virus Cases Using COVID-QF Framework and figure 11 and figure 12 show the recovered cases and deaths respectively in all world. Figure 13 shows Comparison between CORONA Virus total, Recovered, Active and Deaths Cases Using COVID-QF Framework. Figure 14 and Figure 15 show the statistics for the verified deaths rate and detail by comparing cases in age parameter where figure 16 shows average cases for sex of Coronavirus Deaths.

This study's statistics represents the estimated mortality rate and all details found in this study are provisional and prone to change. This data covers instances of New York City citizens and immigrants staying in buildings in New York City.

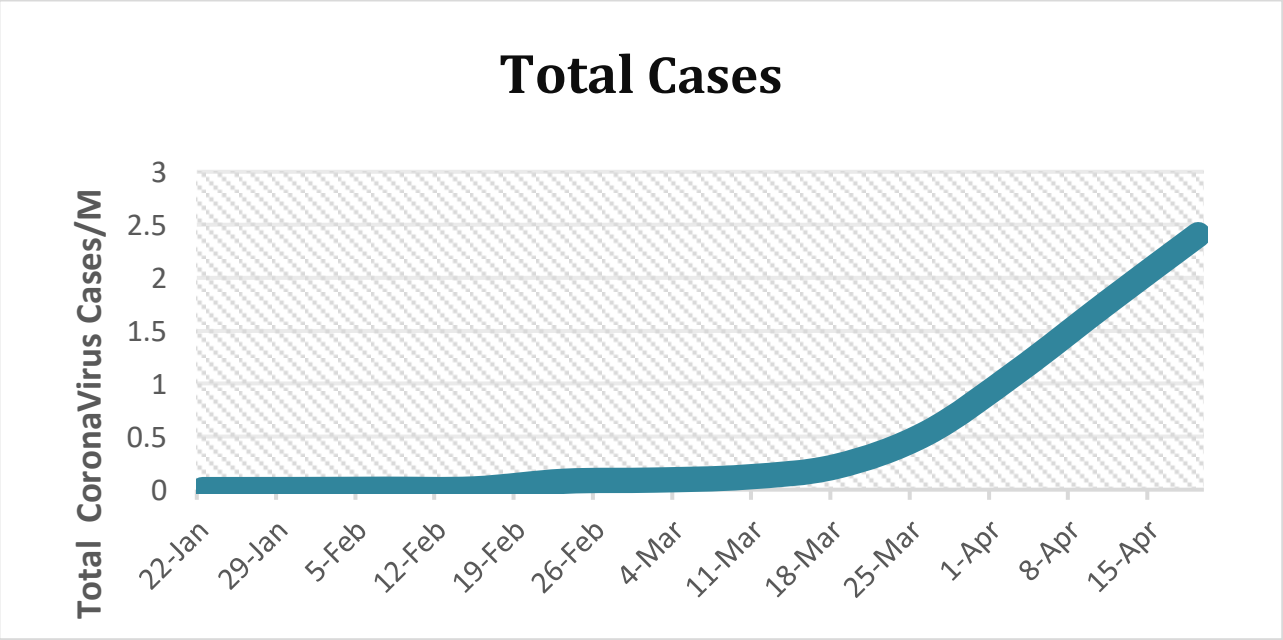

**Figure. 9:** Retrieving Total CORONA Virus Cases Using COVID-QF Framework

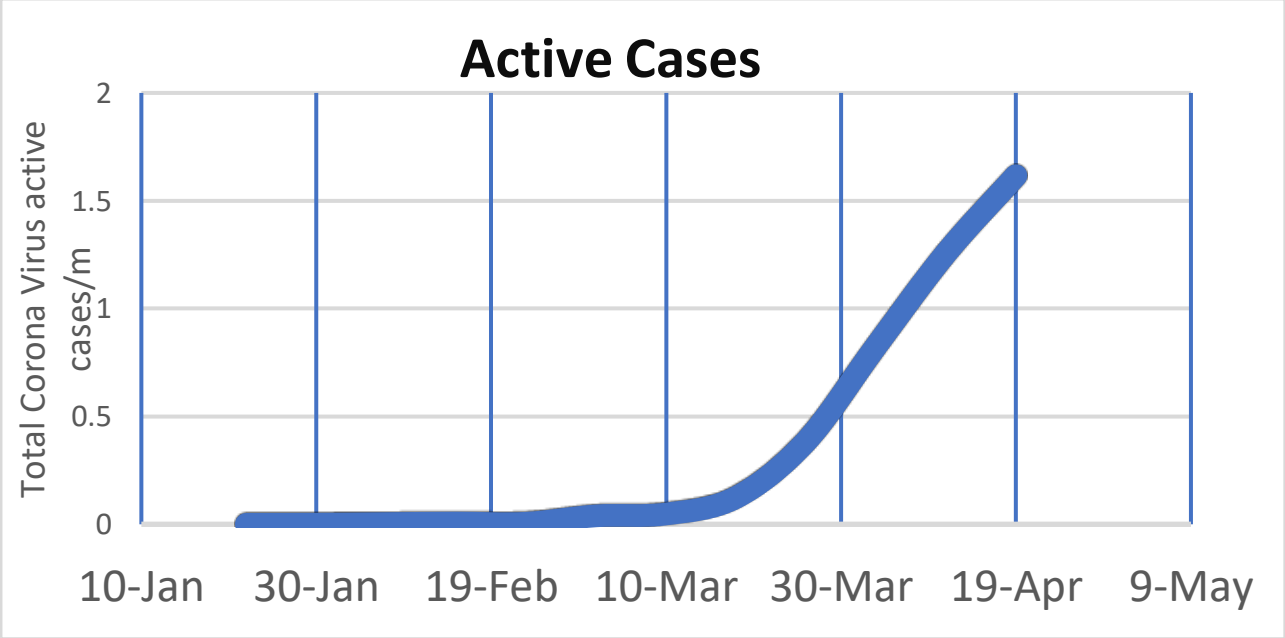

**Figure. 10:** Retrieving Active Corona Virus Cases Using COVID-QF Framework

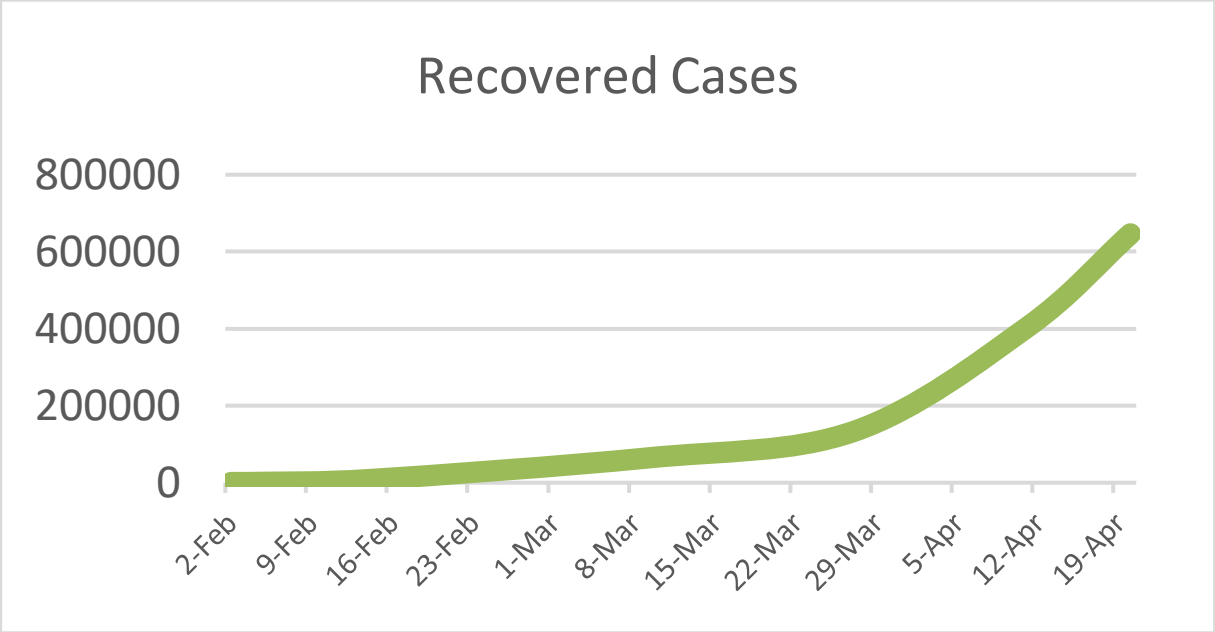

**Figure. 11:** Retrieving CORONA Virus Recovered Cases Using COVID-QF Framework

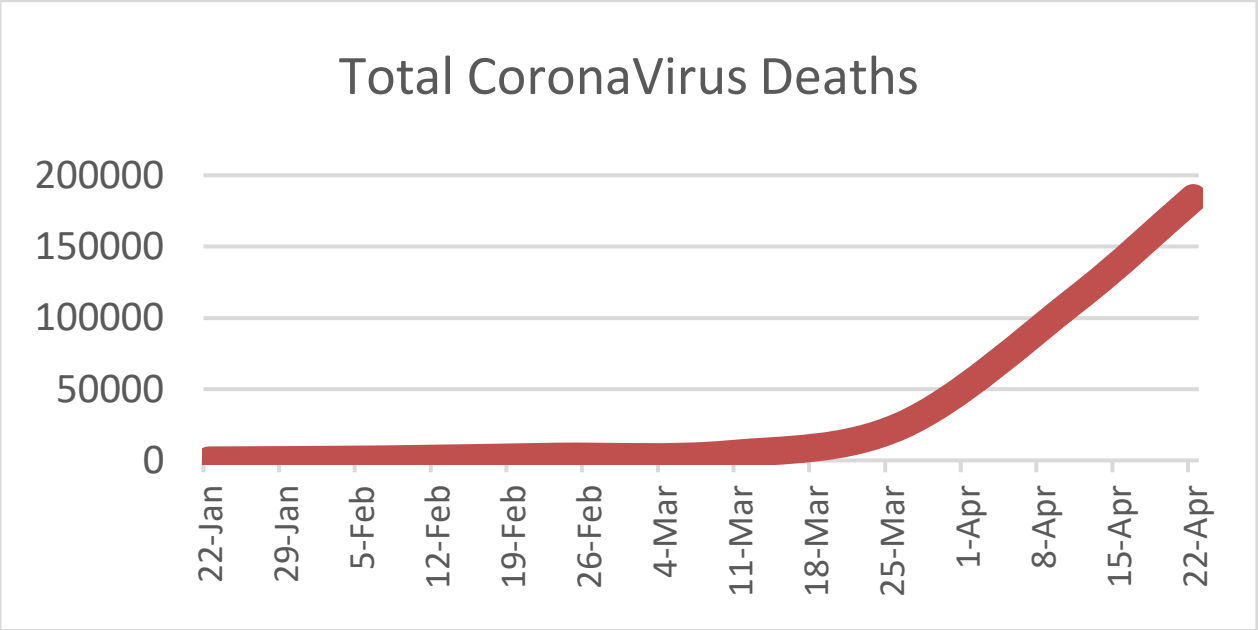

**Figure. 12:** Retrieving CORONA Virus Deaths Using COVID-QF Framework

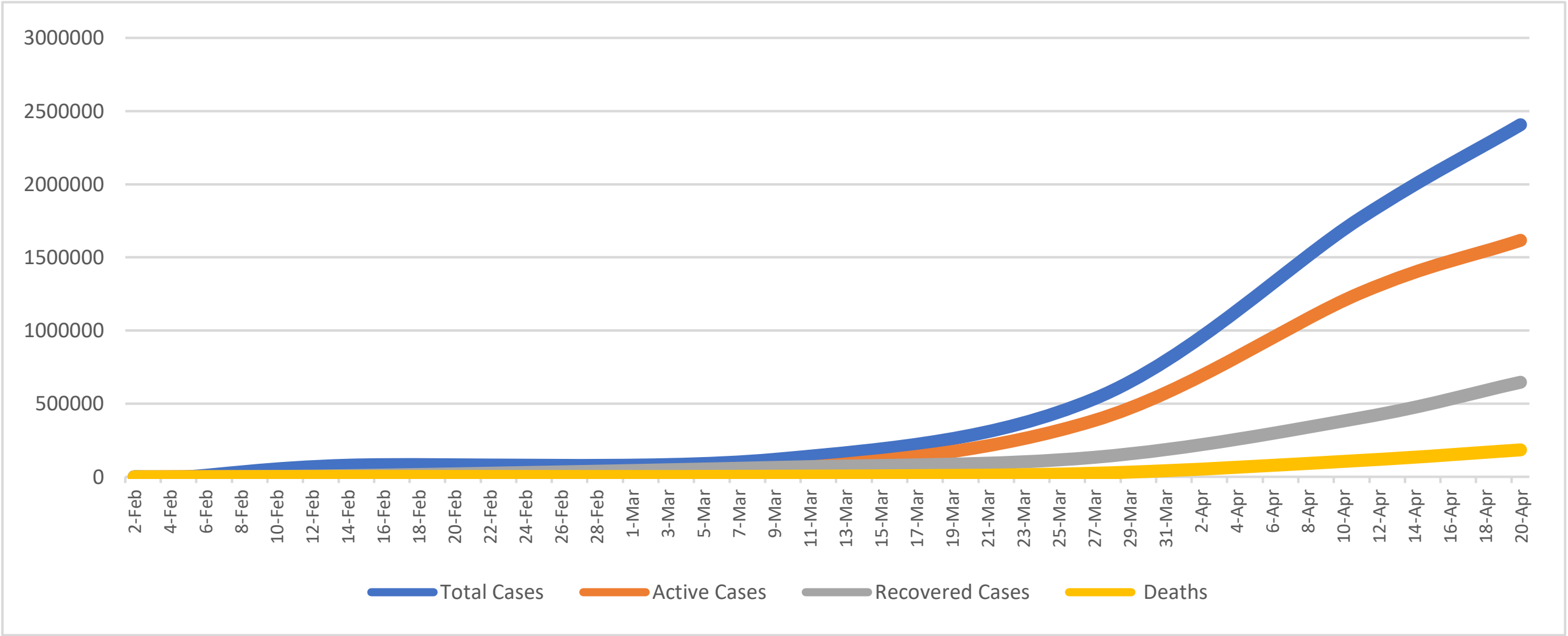

**Figure. 13**. Comparison between CORONA Virus total, Recovered, Active and Deaths cases Using COVID-QF Framework

.

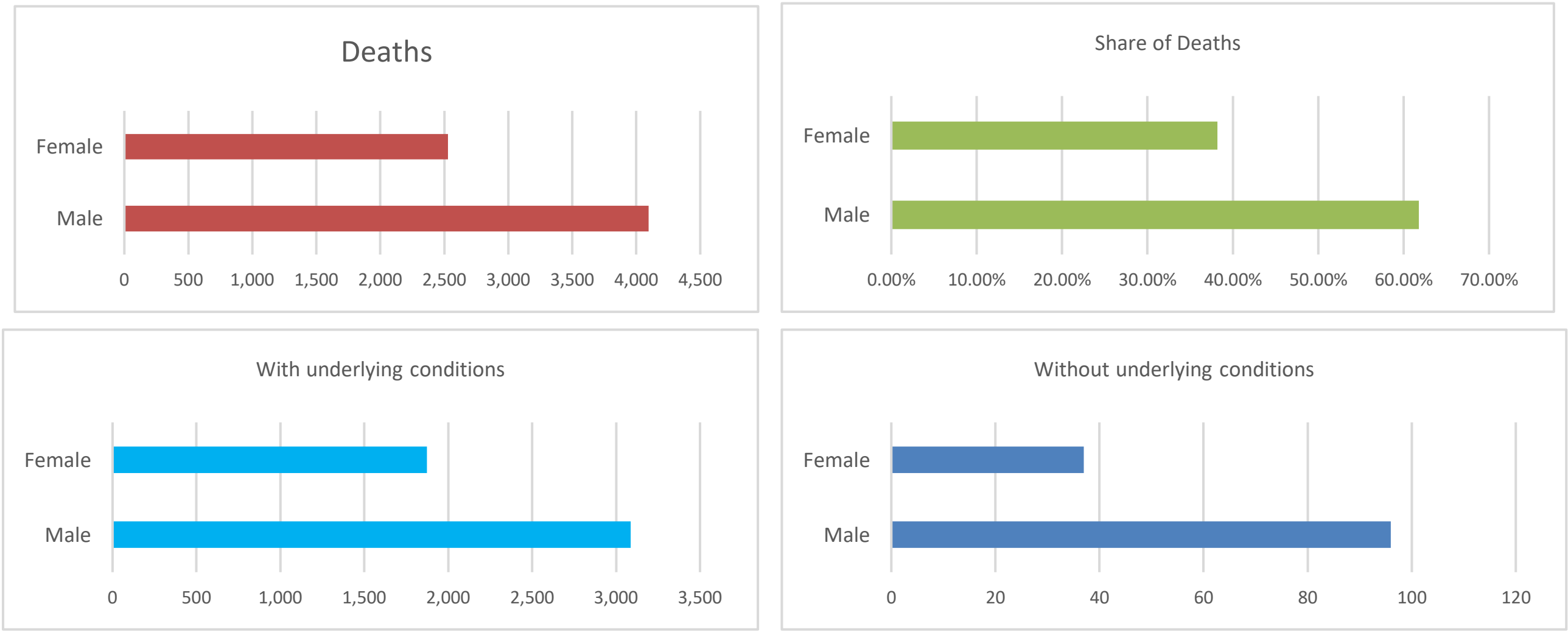

**Figure. 14:** Retrieving CORONA Virus Deaths Using COVID-QF Framework

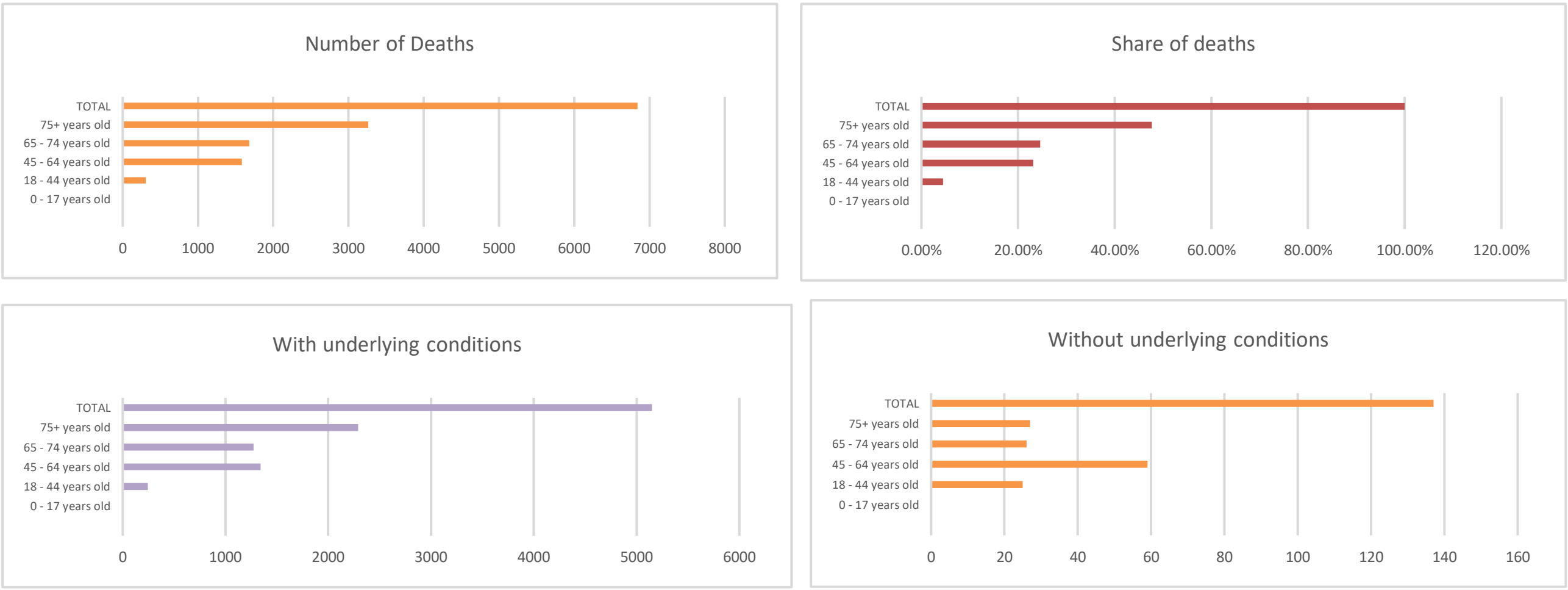

**Figure. 15**: Retrieving Average Cases for Age of Coronavirus Deaths Using COVID-QF Framework

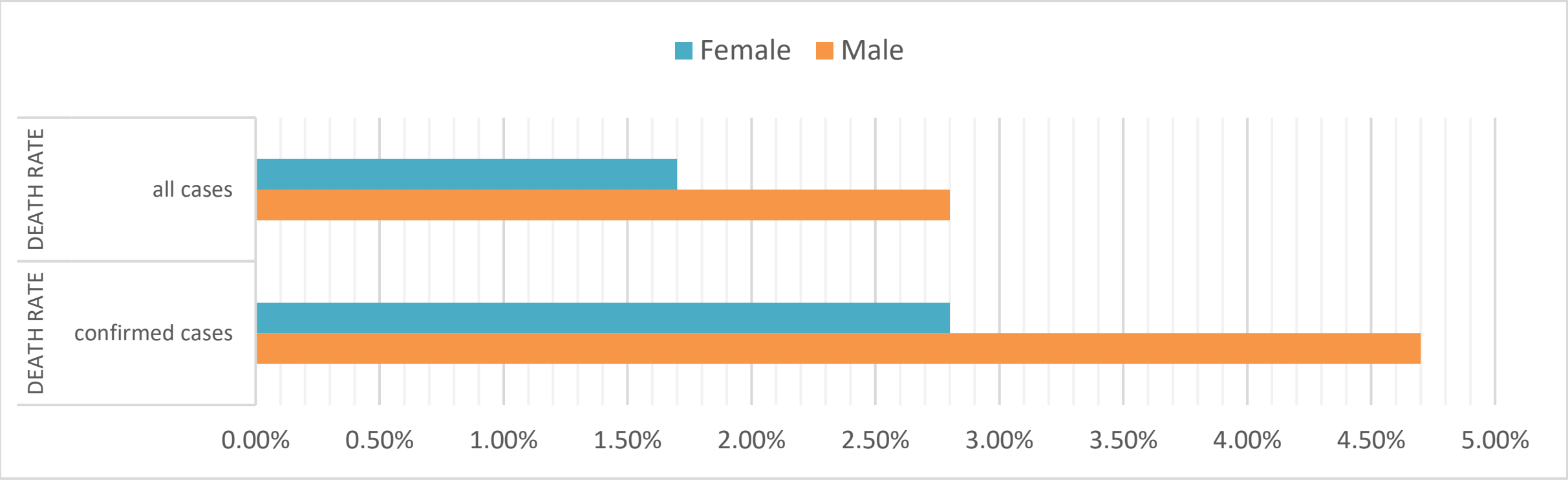

**Figure. 16**: Retrieving Average Cases for Sex of Coronavirus Deaths Using COVID-QF Framework

## 5- Conclusion and Future work

There is a request for additional efforts to perform complex queries over different COVID-19 data sets and different sources. This paper introduced a framework to handle complex query for COVID-19 datasets named COVID-QF. Whereas, the corona takes place in huge numbers in the world, with which only bigdata application and the work of NOSQL databases are suitable. A framework valid for analysing small numbers of data via SQL or large numbers of data via NoSQL. The size of the COVID-19 data collected for each country may reach a large volume over time. The SQL form may be insufficient to handle this size, thus in this case the NOSQL databases should be used as certain. COVID-QF (comprehensive storage COVID-19 data framework utilizing Apache Spark and HDFS), for indexing and processing broad files, is suggested for use in the current research paper. This framework consists of three layers. The first of which is responsible for This layer checks the dataset size, large or small then check for the suitable dataset engine that id matched with the user query sentences. In second layer the system sends user queries to a processing layer containing Hadoop HDFS to store data, the k-aggregation algorithm with MapReduce. The last layer either works with the SQL engine or the selected NOSQL Engine to do the job required. It should be noted that firstly, a vector holding the names of the SQL and NOSQL Engine is created to help in defining the database engine matched with the user query. This paper proposes a time model for calculating time cost and therefore it used Sharding technology with Mongo database queries to segment data and reduce the time used to query.